# Awakening as a Trait-Level Transformation of Consciousness

*Riku Ihalainen[1,2*], Victor Oswald[1,2*], Gulshan Kumar[1,2], Genevieve Major[1,2], Balachundhar Subramaniam[5], Steven Laureys[1,2,3,4,5], Prejaas K.B. Tewarie[1,2,6]*

[1] Faculty of Medicine, Laval University, Québec, Canada
[2] Joint International Research Unit on Neuroplasticity, CERVO Brain Research Centre, Université Laval, Quebec City, QC, Canada
[3] International Brain and Consciousness Institute, Hangzhou Normal University, Alibaba Business School and Hangzhou Institute of Medicine, Chinese Academy of Sciences, Hangzhou, China
[4] GIGA Consciousness Research Unit and HEC Liège Management School, University of Liège, Liège, Belgium
[5] Beth Israel Deaconess Medical Center, Harvard Medical School, Boston, MA, USA
[6] School of Physics, University of Nottingham, UK

[*]Equally contributed

Corresponding authors:

Riku Ihalainen CERVO Research Institute, Laval University, Québec, Canada 2525 Chemin de la Canardière, Québec, QC G1J 2G3, Canada, riku-johannes.ihalainen.1@ulaval.ca

Victor Oswald– CERVO Research Institute, Laval University, Québec, Canada 2525 Chemin de la Canardière, Québec, QC G1J 2G3, Canada, victor.oswald@umontreal.ca

# Abstract

Awakening is reported across contemplative traditions yet remains poorly defined in science, where it is often conflated with transient states induced by meditation or psychedelics, obscuring its significance for theories of consciousness and cognitive architecture. We treat awakening as a functional construct capturing convergence across Hindu and Buddhist sources, phenomenological and psychometric findings, and neuroimaging evidence. Synthesizing these literatures, we find convergent evidence for reorganization across dimensions including self-related processing, affective reactivity, and temporal structure, while exposing limits of paradigms for detecting trait-level change. We argue that awakening is best understood not as a transient state but as a trait-level reorganization of experience, centered on a reduction in identification with mental contents rather than the acquisition of new content. To account for this contentless transformation, we propose the enactive-dereification model (EDM), which treats training and awakening as graded differences along a dereification gradient, from a reified to a relational mode of experience. We offer an operational definition that distinguishes awakening from state-dependent experiences and psychopathology, emphasizing dynamical flexibility and stabilization of a dereified mode. The framework yields falsifiable predictions linking independently assessed trait transformation to large-scale neural dynamics and provides a basis for distinguishing adaptive contemplative change from pathological ego-dissolution.

**Keywords:** Awakening, Contemplative tradition, Dereification, Enactive Cognition, Nondual Awareness, Cognitive Restructuring

# 1. Introduction

"Awakening" is a term used across contemplative traditions to describe reports of a profound shift in subjective experience, commonly involving changes in self-experience and perception and widely regarded as meaningful beyond any single episode of realization. The term carries no single, agreed-upon referent: traditions differ, at times substantially, in how they characterize the phenomenon, its conditions, and its significance, and, as detailed below, no consensus exists within or across them as to what awakening ultimately consists of. Rather than adjudicate between these accounts, we examine their points of convergence and, informed by phenomenological and neuroscientific evidence, define a working construct — which we continue to call awakening, for continuity with this literature — intended to capture a structural core of what is reported rather than to settle what any given tradition takes awakening to be.

Despite, or perhaps because of, this lack of consensus, awakening has regularly been treated as either doctrinally specific or empirically intractable (Davis & Vago, 2013) and has been largely excluded from theoretical accounts of cognition and consciousness. Yet it presents a challenge that warrants scientific attention: how can a transformation that is not principally defined by the addition of a new category of sensory or mental content nonetheless produce an enduring reorganization of how such content is perceived, interpreted, and related to? Answering this requires moving beyond state-based descriptions toward frameworks capable of capturing large-scale structural change in cognitive organization.

Meeting that requirement depends, first, on distinguishing the contents of experience from the structure through which they are experienced. By content we mean what appears in experience — thoughts, emotions, bodily sensations, perceptual objects, and actions, together with their first-order qualities; after awakening, a sound is still heard as a sound, an emotion still felt as an emotion. By structure we mean the relations among contents and the way their experiencing is organized: whether contents are appropriated as "me" or "mine," the articulation of experience into subject and object, the distribution of salience across the experiential field, and the temporal framing of what appears. Awakening, we argue, involves no addition of new experiential content but a reorganization of this structure. The task, then, is not to explain the emergence of new content but to characterize how such a structural reorganization comes about and is sustained over time.

The theoretical resources for characterizing such reorganization have been developing for three decades. Classical computational accounts of cognition, centered on internal symbol manipulation, have progressively given way to frameworks emphasizing the situated, embodied, and relational nature of cognitive processes. The 4E paradigm — enactive, embodied, embedded, and extended — conceptualizes cognition as emerging from continuous interaction between brain, body, environment, society, and, increasingly, technology (Alexander, 2025), so that cognition is no longer a disembodied information processor but a dynamic system shaped by action, context, and lived engagement with the world (Carney, 2020; Newen et al., 2018). Contemplative science has been an early and underacknowledged driver of this shift: Varela, Thompson, and Rosch's (1991) argument that cognition is embodied and that first-person experience is legitimate scientific evidence laid the groundwork for treating meditative practice as experimental access to structured transformations in attention, perception, and self-related processing (Dahl et al., 2015; Kabat-Zinn, 2003; Lutz et al., 2008).

We adopt this enactive, dynamical framing deliberately. A content-centered account, which explains change through new representations, associations, or consolidated memories, cannot capture a transformation that introduces no new content; an enactive account instead locates change in the organization of the system itself—in patterns of coupling, stability, and transition—and is therefore better suited to the structural reorganization at issue here.

Within this framing, awakening is among the phenomena most directly relevant to contemplative science, yet it remains strikingly understudied (though see Sections 4 and 5). This synthesis is not a systematic review: we selected scriptural, phenomenological, psychometric, and neuroimaging sources purposively, by their relevance to three questions — first, how contemplative traditions distinguish awakening from transient altered or absorptive states; second, whether contemporary phenomenological and psychometric studies support stable, multidimensional changes in self-related processing, affect, and the temporal structure of experience; and third, whether neural evidence identifies candidate mechanisms of state-to-trait reorganization. We therefore do not claim exhaustive coverage of either the contemplative or the meditation-neuroscience literature, and we read classical sources functionally and as hypothesis-generating rather than doctrinally comprehensive.

The paper proceeds in four steps. We first examine contemplative models of awakening across South Asian traditions, highlighting their convergences and divergences (Section 2). We then define awakening as a trait-level reorganization of the global structure of experience (Section 3). Next, we synthesize phenomenological reports and emerging neural evidence to characterize it as a stable, trait-level phenomenon (Sections 4–5). Finally, we develop an integrative account — the enactive-dereification model (EDM) — in which awakening is understood as a reorganization of the dynamical structure of relational information processing (Section 6). Within this framework, awakening is treated not as an anomalous or ineffable event but as a lawful transformation in how cognitive systems regulate, integrate, and enact experience over time.

# 2. Contemplative Models of Awakening

The purpose of this section is to frame and examine contemplative models of awakening through a functional framework that bridges them with cognitive science, rather than to provide a full critical analysis of awakening as described in the traditional scriptures. To that end, we map awakening in terms of initial conditions, transitional dynamics (characterized typically by cessation and recognition), and stable trait-like outcomes. An extensive search and characterization of traditional literature is beyond the scope of this paper; instead, we draw on what we consider key scriptural examples and their interpretation. While awakening is described across traditions and cultures, classical South Asian contemplative traditions provide long-standing, textually systematic, and internally coherent philosophical accounts of awakening as a stable transformation of consciousness rather than a transient experiential state, as argued below. This focus also reflects the fact that the meditation-neuroscience literature we draw on in Section 5 is itself weighted toward Buddhist(-derived) practices, particularly mindfulness, relative to other contemplative traditions.

## 2.1 Classical Hindu Models of Awakening

The section draws primarily on three classical Indian philosophical schools: *Yoga*, *Sāṃkhya*, and *Advaita Vedānta*.

### *2.1.1 Yoga and Samhkya*

Yoga and Samkhya are interlinked philosophical schools that together articulate one of the earliest formal psychological models, in which mental activity is treated as an object of observation rather than the subject of experience (Bryant, 2009; Feuerstein, 1979; Flood, 1996). Central to *Sāṃkhya* is the question of suffering, which is understood as the consequence of ignorance (*avidyā*), the foundational misidentification of awareness with cognitive activity. The foundational aphorism defines yoga as "the stilling of the changing states of mind" (YS 1.2), while the immediately following aphorism (YS 1.3) specifies that "awareness then abides in its own nature" (Bryant, 2009; Saraswati, 2013). This attainment, which the tradition treats as its account of awakening, is thus not defined by any particular experiential content but by a persistent disidentification of awareness from mental activity (Kelly & Whicher, 2015).

Yoga postulates that not-knowing can be understood as a hierarchy of afflictive processes (ignorance, subject-identification, attraction, aversion, and the drive toward self-continuity). Ignorance itself is the root, the foundational misidentification of awareness with cognitive activity. From it arises a dynamic self-model in which awareness takes itself to be identical with thinking, feeling, and the sense of continuous agency. From that self-model follow automatic patterns of attraction toward what confirms it and aversion toward what threatens it. These consolidate into a systemic drive toward self-continuity that is not psychological anxiety but a structural tendency of the organization to resist any dissolution of its own coherence. The hierarchy is causally closed: each layer generates the next, and the deepest layer reinforces the root.

The contemplative method described by Yoga works along the reverse order of the afflictive hierarchy, progressively destabilizing the automatic coupling between cognitive content and awareness. At each stage, awareness partially withdraws from identification with a layer of the hierarchy, a process the tradition describes as a progressive uncoloring: cognitive processes continue, but awareness ceases to take their color as its own. A critical feature of the *Yoga* framework is the explicit differentiation between *samādhi* (transient states of attentional absorption) and *kaivalya* (a trait characterized by the disidentification of awareness from mental processes). While samadhi refers to refined but reversible experiential states, *kaivalya* denotes the final aim of Yoga: the isolation or liberation of awareness from identification with the psychophysical complex (Saraswati, 2013; Vasu, 1914). Deep absorptive states (*samādhi*) represent the transitional moment in this process, in which self-identification temporarily ceases and awareness rests in itself without appropriating content. Classical commentators emphasize that even the most subtle absorptive states remain conditioned, whereas awakening (*kaivalya*) reflects a trait-like reconfiguration of the relationship between awareness and cognition that persists beyond meditative practice (Muktibodhananda, 2008; Whicher, 1998; Wilber et al., 1986).

*2.1.2 Advaita Vedānta: State-independent transition of self-related processing*

Advaita Vedānta offers one of the most systematic accounts of awakening as a state-independent transformation of self-related processing rather than a transient altered state (Deutsch, 1969; Mayeda, 1992). Central to this framework is *adhyāsa*, the mutual superimposition of self and non-self upon each other. Awareness is misattributed to the properties of mental and perceptual processes, such as limitation and change, while those processes are simultaneously appropriated as "I" or "mine" (Śaṅkara, trans. Madhavananda, 1921; Radhakrishnan, 1953). This bidirectionality matters: *adhyāsa* is not a one-directional attribution error but a reflexive coupling in which awareness and the psychophysical complex mutually define each other's apparent properties. The phenomenon is experientially familiar: we say "I am tired" rather than "tiredness is occurring" precisely because awareness has borrowed the properties of the body-mind, while the body-mind has borrowed the apparent luminosity of awareness. From a cognitive perspective, this resembles a persistent self-content coupling mechanism through which representational processes become bound into a stable subject-object structure organized around ownership, agency, and self-referential continuity. Unlike *Yoga* and *Sāṃkhya*, *Advaita Vedanta* targets not only mental activity or the awareness-cognition distinction, but also the superimpositional process through which subject, object, ownership, and agency are constituted together as a unified structure.

Advaita destabilizes this coupling through the *dṛg-dṛśya viveka* analysis, a progressive discrimination between the observing awareness and the observed. Body, sensations, thoughts, emotions, and even the autobiographical self-image can each be made into objects of awareness. The logical force of this observation is immediate: whatever can be observed cannot itself be the observer. If the body appears to awareness, the body is not awareness. If a thought appears to awareness, the thought is not awareness. If the sense of being a self appears to awareness, that sense is not awareness either. The analysis proceeds recursively through every candidate for selfhood until no object remains that can credibly claim to be the subject. Crucially, the analysis does not terminate by arriving at a purer or more refined witness. Rather, it dissolves the very subject-object axis through which witnessing is organized, leaving what Advaita describes as *nirviśeṣa cit*, awareness that is self-luminous and without internal differentiation, knowing without a knower rather than a blank or empty state (Deutsch, 1969; Nikhilananda, 1949). Awakening here does not involve acquiring new experiential content, in the sense specified in the Introduction, but the removal of a systematic misidentification that was previously organizing experience, while perception, cognition, and behavior continue to function (Fort, 1998). Hence, awakening is not even a witness in the technical sense, because witnessing still implies a relational structure. It is better understood as the nondual ground that neither arises nor ceases, within which the apparent movement of states occurs (Easwaran, 2007; Nikhilananda, 1952; Radhakrishnan, 1953).

### *2.1.3 Distinguishing Awakening from Psychopathology*

Classical sources are explicit that liberation is distinct from pathological detachment. In Yoga, misidentification is treated as a cognitive error (*avidyā*), while awakening is marked by clarity (*viveka*) and stability rather than confusion (Bryant, 2009; Saraswati, 2013; Whicher, 1998). Similarly, *Sāṃkhya* and Advaita *Vedānta* make the same distinction, emphasizing preserved discriminative insight and functional engagement even as false identification dissolves (Larson, 1987; Phillips, 2009; Radhakrishnan, 1953; Śaṅkara, trans. Madhavananda, 1921).

From a clinical perspective, this distinction is decisive. Conditions such as depersonalization-derealization disorder involve distressing detachment from self or world, emotional numbing, and functional impairment, while reality testing is typically preserved (American Psychiatric Association, 2013; Sierra, 2009; Simeon & Abugel, 2006). By contrast, awakening as described across these traditions is marked by increased stability, enhanced discriminative awareness, preserved functional integration, reduced affective reactivity alongside maintained emotional responsiveness, and peace or equanimity (Lindahl et al., 2017). Recent research suggests that interpretation, context, control, duration, distress, and functional impact all shape whether reduced self-identification is experienced as beneficial, challenging, or clinically concerning (Newberg & Waldman, 2018; Van Dam et al., 2018).

Classical sources also warn against mistaking blissful, absorptive, or extraordinary states of experience for liberation itself (Feuerstein, 1979; Muktibodhananda, 2008; Wilber et al., 1986), anticipating a common contemporary confusion between genuine awakening and altered states arising from pathology, psychedelics, or pharmacological influence (Millière et al., 2018). Just because an experience feels profound or unusual does not, in this view, make it a decisive sign for awakening. The closing aphorism of the Yoga Sūtras (YS IV.34) makes this point with striking clarity. Liberation is defined there not as heightened or enhanced experience but as the exhaustion of self-serving dynamics, the dissolution of the structures that keep the mind oriented around its own agenda, with the emphasis falling entirely on what falls away rather than what is gained (Patañjali, trans. Bryant, 2009; Saraswati, 2013). Awakening, on this view, should not be identified with altered arousal, reward, or sensory vividness, which may accompany contemplative experience without serving as its measure (Berkovich-Ohana et al., 2013; Bronkhorst, 1993; Josipovic, 2014).

## 2.2 Buddhist Models of Awakening

In this section we consider Theravada and Mahayana Buddhism, and although Vajrayana is doctrinally a branch of Mahayana, we treat it separately here. For Theravada, we consider the Samyutta Nikāya, and particularly a key set of suttas, SN 22, SN 43, and SN 56 (Bodhi, 2000), as the basis from which we can derive an understanding of how awakening is characterized under this tradition. For Mahayana, we mainly focus on the Heart Sutra (Conze, 2001), and for Vajrayana we focus on Dzogchen and consider the three statements of Garab Dorje as a key literature piece (Garab Dorje, 1996).

### *2.2.1 Initial Conditions – Root Causes of Suffering*

In Theravada, the starting condition is described as fundamental ignorance of the four noble truths – not knowing and understanding suffering, its origin, its cessation, and the path (SN 56.17). This manifests as identification with and clinging to the five constituents of experience (khandhas) **-** form, feeling, perception, mental formations, and consciousness that is taking them as 'self,' 'mine,' or 'what I am.' SN 22 emphasizes repeatedly relinquishing the view of these constituents as "mine" or as "what I am", with the cessation of such identification marking liberation. Hence, the starting condition can be succinctly characterized as the *presence of this misidentification*.

Mahayana tradition describes the starting condition as a fundamental misapprehension of 'emptiness' or sunyata, that is, as the false belief that phenomena possess an inherent, often independent, existence. This reifies the phenomena, including 'emptiness' itself, as possessing an intrinsic nature, a stance that the tradition rejects, since all phenomena arise dependently. Taking anything as independently real or as a ground for reality is rejected as a mistake. The Heart Sutra makes this explicit when it states, "form is emptiness, emptiness is form" (Conze, 2001), often interpreted as emphasizing precisely that phenomena lack inherent and independent existence. Accordingly, the initial condition can be characterized by conceptual and ontological reification, as well as attachment to views and experiences.

In Vajrayana, particularly in Dzogchen as articulated by Garab Dorje, the starting condition is described as failure to recognize the nature of mind (rigpa) (Garab Dorje, 1996). Ordinary experiences are dominated by distraction and grasping to and identification with thoughts, perceptions, and even meditative states. This ignorance is framed as non-recognition and crucially not as a lack of development or skill. The starting condition is thus characterized as *misidentification with arising phenomena rather than recognition of awareness itself*, along with habitual re-engagement with conceptual elaboration.

As such, all three traditions describe the starting condition via negation: as misapprehension or misidentification, rather than lack of e.g. experience, skill, awareness, or capacity. Therefore, the solution cannot be adding anything but letting go of the confusion (cessation of misapprehension and misidentification). This does not imply that the path is purely subtractive: some Buddhist models describe extensive cultivation of attention, ethics, compassion, wisdom, and meditative stability as conditions through which recognition becomes possible and progressively embodied (Rinpoche, 2002)

*2.2.2 Transition*

Across the three traditions, the transitional dynamics of awakening are understood as cessation, deconstruction, or recognition, albeit with differing emphasis. In Theravada, the emphasis is on progressive cessation of clinging to the five constituents of experience, grounded in insight (vipassana). This involves repeated observation of the constituents without identification or appropriation.

Mahayana frames this transition as the collapse of reification, obtained similarly via insight. Here, the insight, which is insight into emptiness, involves seeing through the conceptualization that posits inherent existence and leads to cessation of grasping at phenomena and concepts, e.g. views, practices, and even attainment itself.

In Vajrayana (particularly in Dzogchen), the emphasis is on direct recognition of the nature of mind (awareness) and integration of that recognition. The transition dynamics consists of interrupting the typical, habitual identification with arising phenomena and of repeated re-recognition if and when distraction occurs. Thus, whether framed as insight into non-self, emptiness, or the nature of mind, the transition across these traditions functions by weakening and eventually dissolving misidentification, rather than by adding new cognitive or phenomenological capacities.

*2.2.3 Stable outcome – Nothing has Changed but Everything is Different*

In all three traditions, the stable outcome is characterized by continuous cessation or recognition; only the emphasis varies. Theravada characterizes the cessation of ignorance and clinging as the stable awakening. SN 22 and SN 43 highlight that the constituents continue to arise; what has changed is the relinquishment of them as "mine" or "part of myself". Hence, awakening is considered *a sustained shift in how experience is related to (non-identification, reduced reactivity)* and not as maintaining an (altered) experiential state.

Mahayana describes this as continuous non-abiding wisdom that is grounded in insight into emptiness. Following this insight, experiences continue to arise, but now without attachment to any views, experiences, or concepts. The Heart Sutra's rejection of attainment ("no attainment and no non-attainment") reflects this trait-like condition: awakening is not sustained by holding a realization, but by *the absence of reification*.

Finally, in Vajrayana (Dzogchen), this stable outcome is described as *continuous recognition of the nature of the mind* (while initially emerging intermittently); a nature characterized by effortless presence and non-identification with mental content across context. Again, thoughts, emotions, and perceptions continue to arise but without resulting in distraction and identification with them.

**Box 1. Minimal operational definition.**

Awakening is defined here as a durable, context-independent reorganization of the global structure of experience, characterized by reduced self–content coupling, altered affective reactivity, altered temporal organization of experience, and diminished dominance of self-referential evaluation. It is not defined by the presence of a specific altered state, doctrinal commitment, or metaphysical interpretation.

Therefore, stable awakening in the Buddhist models is not defined by a particular phenomenological state, but rather as a continuous, persistent alteration in how one relates to experience, how one acts, and how experience itself is structured.[1] Ordinary cognitive and perceptual processes continue ('nothing has changed...'), but misidentification, reification, and clinging have been dropped ('... but everything is different').

[1] The content-independence claims reviewed in this section concern the transition itself, not necessarily the training that precedes it. Contemplative training often proceeds through the deliberate construction of new content, visualization, conceptual reframing, as a means toward that transition, and this varies considerably across traditions and practices. This is not in tension with the account above: training may operate through content while the transition it leads to does not, on the traditions' own terms, consist in acquiring further content. Such content is often deployed instrumentally, as scaffolding used to dismantle deeper patterns of identification, rather than as an end in itself.

Figure 1 summarizes this convergent structure across all five traditions discussed above, alongside the corresponding Hindu models of Yoga, Sāṃkhya, and Advaita Vedānta.

**Figure 1.**

*Convergent structure of awakening across five classical South Asian contemplative traditions.*

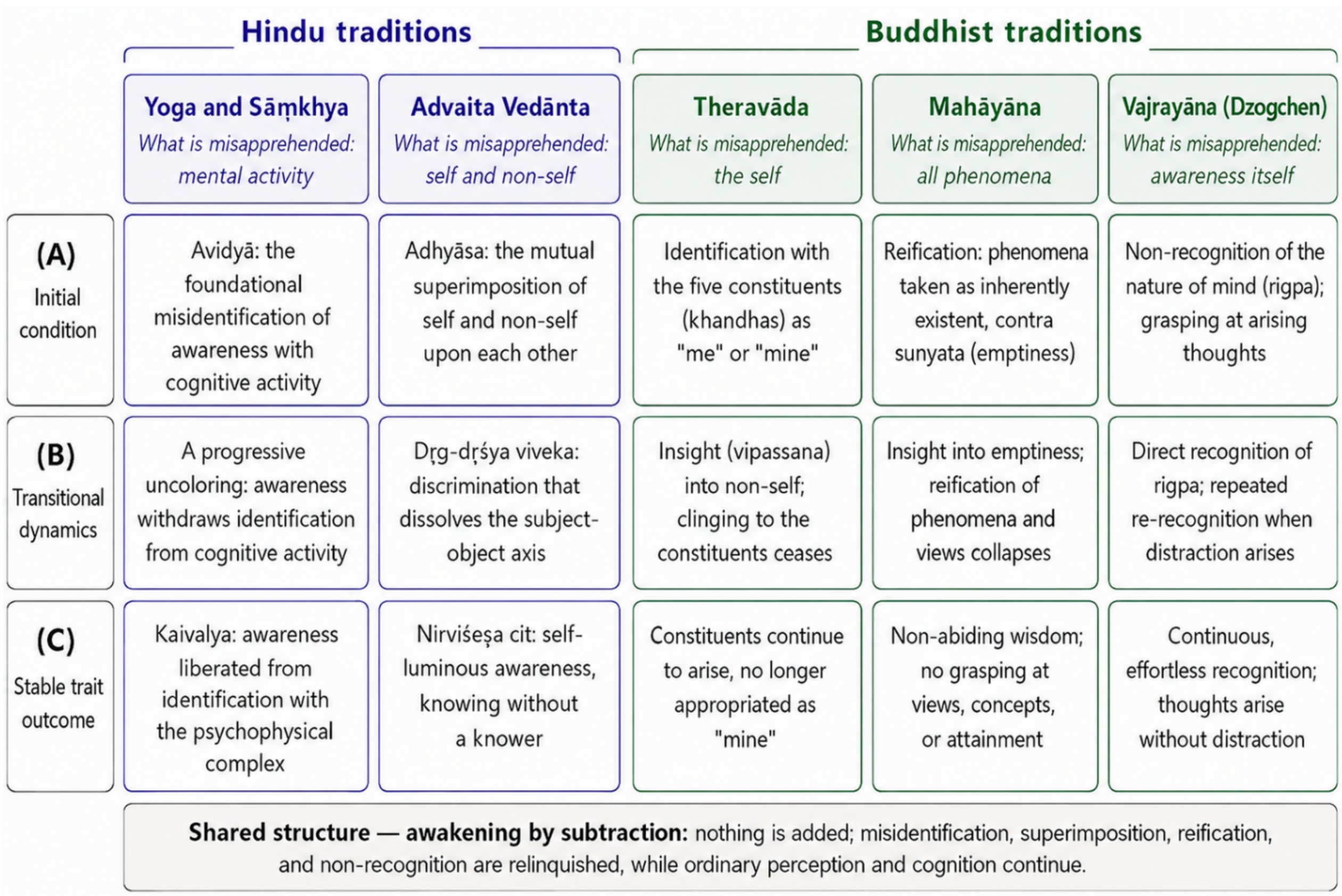

*Note.* For each tradition (Yoga and Sāṃkhya, Advaita Vedānta, Theravāda, Mahāyāna, and Vajrayāna/Dzogchen), the figure summarizes the initial condition identified as the root of suffering or misapprehension, the transitional dynamics through which it is addressed, and the resulting stable trait-level outcome, as characterized in the source texts discussed in this section. Despite differing terminology and targets, all five traditions characterize awakening via negation, as the relinquishment of a specific misidentification,

superimposition, reification, or non-recognition, rather than the acquisition of new experiential content.

### 2.3 Towards an Operational Definition

Characterizing stable awakening via negation, as something subtractive rather than additional, and as non-phenomenal yet productive of persistent first-person changes, places it outside the standard descriptive vocabulary of cognitive science. Rather than presenting as a state or variable in the usual sense, it is better approached as a pattern of organization and identification. From the perspective of cognitive science, this suggests that awakening needs to be approached not as a state contrast, but as a stable reconfiguration of self-related processing and system-level dynamics. While the contemplative models converge on this idea, on awakening as a persistent cessation of misidentification, translating this into an empirically tractable framework requires an explicit operational definition. In the following section, we develop an operational definition that captures this structural insight in terms amenable to empirical investigation.

## 3 Defining Awakening: A Trait-Level Construct

### 3.1 State or Trait? From Transient Experiences to Enduring Transformation

Contemplative traditions frequently describe a range of altered experiential states, such as nondual awareness (NDA), cessation, pure awareness, or absorptive states, that are often taken as markers of advanced practice or awakening. However, these states raise an important conceptual issue for cognitive science. First, such states are typically transient and unstable: individuals do not remain in them continuously, and in many cases, they appear functionally incompatible with everyday cognition and action. Second, even when these states are reported as profound or transformative, there is no clear mechanism explaining precisely how transient experiential episodes contribute to enduring, trait-level changes. This distinction highlights a critical gap: while altered states may act as catalysts or perturbations, they do not, in themselves, explain the emergence of what contemplative traditions describe as awakening. Awakening is not characterized by the maintenance of a particular experiential state, but rather by a stable transformation in how experience is structured and related over time.

## 3.2 The Enactive Challenge

Within an enactive framework, cognition is intrinsically relational: it is always consciousness of something, whether thoughts, emotions, bodily sensations, or environmental affordances. Experience arises through the organism's ongoing, embodied engagement with the world. Cognition, in this view, is not the manipulation of internal representations, but a process of sense-making enacted through dynamic organism–environment coupling. At the same time, cognitive science typically explains enduring change through mechanisms such as learning, memory consolidation, or reinforcement of representational structures. These mechanisms presuppose modifications in content, new information, or new associations.

Yet empirical and phenomenological evidence converges on a different picture: awakening is associated with durable changes in cognition, affect, and behavior, including: i) reduced self-referential processing, ii) altered emotional reactivity, iii) increased psychological flexibility, iv) stable shifts in values and meaning-making. This leads to a seemingly significant explanatory tension: How can intensive training produce a lasting reorganization of the cognitive system when the transformation is not principally defined by the acquisition of new informational or experiential content?

## 3.3 Beyond Content-Centered Models

At its core, this seeming paradox exposes a limitation in content-centered models of cognition. If cognitive change is assumed to arise primarily through modifications in representational content or discrete experiential states, awakening becomes difficult to explain. The phenomenon instead suggests that transformation can be characterized in terms of the global organizational principles governing how contents are engaged, integrated, and identified with over time. Rather than introducing new contents, awakening appears to involve a reconfiguration of that organization: a shift in explanatory focus from local, state-dependent changes to system-level dynamics.

## 3.4 Defining Awakening as Trait-Level Reorganization

For present purposes, awakening can be defined as a durable reorganization of the global structure of experience that does not depend on the acquisition of new experiential content. Structural change, in this context, refers not to the emergence of new content, but to alterations in how such content is organized and related to, often following extensive training in deconstructing misidentification.

Several candidate dimensions of this reorganization can be identified:

1. <u>Self–content coupling:</u> The extent to which thoughts, emotions, and perceptions are automatically appropriated as "me" or "mine," or identified with.

2. <u>Affective reactivity:</u> The default linkage between perception and evaluative or emotional escalation.

3. Temporal organization of experience: The degree to which experience is narratively structured and projected across past and future.

4. Self-referential evaluation: The stability and dominance of ongoing self-monitoring, commentary, and judgment.

It is useful to distinguish two levels of description. At any moment, experience occupies a particular configuration along these dimensions — a state; over time, the recurrence of such states forms a distribution whose shape defines the trait. On this view, awakening is a property of the distribution rather than of any single state: not the attainment of a special momentary experience, but a durable change in which states are visited and how readily the system moves between them. We later characterize this shared axis (Section 6) as a gradient from a reified to a dereified mode of experience, and summarize the resulting reorganization in Figure 2 and Table 1.

Two properties of this reorganization must be distinguished, since the account has sometimes been read as invoking two separate things. Considered as a distribution of states along these dimensions over time, the trait is characterized both by (i) a shift of its mean toward reduced identification and reification—a stabilized, more dereified baseline—and (ii) increased mobility around that mean, that is, a lowered cost of transitions between more and less differentiated modes of engagement. What we later describe as the "stabilization of a dereified mode" corresponds to the first property and what we call "dynamical flexibility" to the second; these are complementary aspects of a single distribution rather than competing definitions. Awakening is therefore not residence in any single mode but a reorganization of how the system is distributed across, and moves between, them.

These dimensions are formulated within a cognitive neuroscience framework and are intended as functional descriptors, rather than direct translations of the contemplative constructs. The construct as defined here does not aim to characterize awakening as understood within any specific tradition, or to engage its broader philosophical and soteriological interpretations, but to identify the functional dimensions that recur across traditions and converge with psychometric and neuroscientific evidence. On this account, awakening involves a shift in the global organization of cognition, raising the question of what dynamical principles could support such trait-level reconfiguration.

# 4. Phenomenological Reports of Awakening-Like Transformations

If awakening is best understood as a trait-level alteration in the structure of experience, as we argue, then it should be observable in lived phenomenology: not as changes in specific experiential contents or states, but as enduring shifts in how experience is structured and related to. The central question, therefore, is not whether extraordinary or peak experiences occur, but whether there is evidence for stable, context-independent reorganization along the dimensions outlined above: self-content coupling, affective reactivity, the temporal organization of experience, and self-referential evaluation. Rather than treating awakening as a doctrinal claim, here, we examine whether subjective reports across contexts reflect these structural features, reflecting a reorganization of how content is experienced. Note that here we refer to such reports as 'awakening-like' rather than 'awakening,' since the cited studies were not designed around the construct as defined here or as characterized within the traditions reviewed above; we return to this distinction at the end of this section.

## 4.1 Contemporary Phenomenological Accounts

Contemporary qualitative and mixed-method research provides converging subjective evidence for awakening-like transformations across religious, secular, and mixed contexts. In a large interview-based study of individuals reported persistent non-symbolic experience (PNSE), reported to arise through diverse pathways including meditation, psychotherapy, psychological crisis, secular inquiry, and spontaneous onset (Martin, 2020). Participants did not describe a uniform phenomenology. Nevertheless, a cluster of recurring features emerged, including reduced self-referential identification, diminished cognitive rumination, altered emotional reactivity, and a stable sense of presence. These features were described as ongoing and continuous rather than episodic, consistent with the possibility of awakening-like trait-level reorganization rather than transient altered states (Martin, 2020).

By contrast, studies explicitly investigating "awakening experiences" often focus on temporary expansions or intensifications of awareness. In a thematic analysis of ninety such reports, Taylor and Egeto-Szabo identified common triggers including psychological turmoil, contact with nature, spiritual practice, and engagement with spiritual literature (Taylor & Egeto-Szabo, 2017). These experiences typically lasted minutes to hours yet frequently produced durable "secondary shifts" in perspective, values, and identity. Although these reports emphasize expansion and intensity during the episode itself, the long-term transformations described tend to concern enduring changes in self-referential stance and reactivity rather than the persistence of heightened perceptual or affective states.

Similarly, large-sample survey research on spontaneous spiritual awakenings reports phenomenological features such as nondual merging, intensified perception, affective elevation, and perceived insight, often accompanied by lasting positive changes in well-being and worldview (Corneille & Luke, 2021). Moreover, although systematic micro-phenomenological investigations of stabilized awakening remain non-existent, such approaches applied to contemplative and nondual states have revealed fine-grained shifts in self-positioning, attentional organization, and temporal unfolding (Dor-Ziderman et al., 2016; Petitmengin, 2006, 2017) - findings that support the view that these transformations involve reorganization in how experience is structured rather than emergence of new perceptual content per se.

Taken together, these findings are consistent with the structural definition proposed in Section 3 and support a distinction between transient experiential episodes and enduring trait-level transformation. While peak episodes may be brief and intermittent, the central issue is whether persistent reorganization of self-related processing and identification is evident. Awakening, as defined here, should not be conflated either with such transient peak experiences or with the depth of meditative absorption achieved during practice.

At the same time, definitional and operational challenges remain. Awakening is a highly underspecified construct unless grounded in specific phenomenological and behavioral criteria (Davis & Vago, 2013), and as Tempone-Wiltshire and Matthews argue, its meaning varies across traditions and may involve non-conceptual or ineffable elements that resist straightforward operationalization (Tempone-Wiltshire & Matthews, 2024). This is part of why we describe the findings in this section, and the psychometric evidence that follows, as 'awakening-like' rather than 'awakening': the underlying construct's variability and resistance to operationalization make direct correspondence claims premature. Nonetheless, a consistent pattern emerges across studies: awakening-like transformations are described not primarily as the acquisition of new beliefs or experiences, but as a persistent shift in reactivity and relational stance toward mental contents. This convergence provides a phenomenological foundation for treating awakening as a trait-level alteration in the structure of experience and raises the question of whether such transformations can be systematically captured by psychometric instruments.

## 4.2 Psychometric Instruments

In contemporary psychological research, awakening-related phenomena are most frequently operationalized under the construct of nondual awareness, defined as diminished subject–object structuring of experience and reduced self-reification. Conceptually, this overlaps substantially with the structural definition of awakening proposed in Section 3, particularly regarding reduced identification with mental contents and altered self-related processing. However, nondual awareness primarily captures reported phenomenological qualities of non-separation and unity, whereas awakening, as defined here, refers more broadly to a stable reorganization of self-related processing and reactivity across contexts. Nondual awareness may therefore represent a central phenomenological dimension of this broader reorganization, but does not, by itself, exhaust the construction itself.

One of the most systematic efforts to operationalize this dimension is the Nondual Awareness Dimensional Assessment (NADA; Hanley et al., 2018). Developed through scale construction and factor-analytic validation, the instrument quantifies dimensions of nondual experience across contemplative and non-contemplative populations. Items probe experiences of diminished subject–object separation, unity, spaciousness, and related phenomenological qualities, and the scale demonstrates internal reliability and meaningful associations with meditation experience and well-being. As such, NADA provides empirical support for the claim that nondual awareness can be assessed as an individual difference variable, though this reflects the scale's reliability and its correlations with related constructs rather than confirmation that it specifically captures nondual awareness as characterized in contemplative accounts. At the same time, its emphasis remains on descriptive experiential features rather than on the underlying dynamics by which mental contents are appropriated, identified with, or regulated over time. This distinction becomes important when attempting to differentiate transient experiential intensification from more enduring alterations in the structure of reactivity and self-related processing.

Beyond NADA, additional instruments attempt to assess more explicitly trait-level dimensions of nondual realization. The Nondual Embodiment Thematic Inventory (NETI; Butlein, 2009) was developed to evaluate qualities of nondual experience and spiritual awakening in daily life, including dimensions such as diminished doership, reduced identification with mental contents, resilience, and capacity to tolerate cognitive dissonance[2]. In principle, these qualities gesture toward relatively stable changes in self-experience rather than momentary unity states. Similarly, the WAKE measure (Kilrea et al., 2023), seeks to classify ongoing shifts in baseline experiential structure rather than transient altered states. Both instruments, therefore, move closer to assessing durable reorganization of self-related processing than NADA. However, important caveats apply: the NETI's original validation remains unpublished, and its primary peer-reviewed study was conducted in the context of a commercial well-being program, raising questions about independence; and both instruments remain grounded in retrospective self-report and culturally specific interpretive frameworks, whereas the construct as defined here aims to capture patterns that recur across cultures and traditions.

---

[2] The primary peer-reviewed psychometric examination of the NETI is Mills et al. (2018), who report high internal consistency (Cronbach's $\alpha = 0.913$) and convergent validity with spirituality measures. However, that study was conducted in the context of a commercial Ayurveda-based well-being program at the Chopra Center for Wellbeing, which represents a potential conflict of interest that limits the independence of the validation evidence.

More generally, questionnaire-based measures rely on retrospective self-report and remain sensitive to interpretive frameworks, linguistic competence, and cultural expectations. They also struggle in distinguishing transient experiential intensification from stabilized trait-level transformation. High scores may reflect temporary absorption states, belief endorsement, or adopted identity narratives, rather than durable reorganization of self-related processing. Indeed, given awakening is defined precisely by stability across contexts and independence from ongoing practice, self-report measures are structurally limited in their capacity to verify it. This motivates the search for convergent evidence at the level of neural and physiological dynamics, to which we now turn.

# 5. Objective Evidence and Neural Correlates

The preceding sections have converged on awakening as a stable, trait-level reorganization of cognitive dynamics across four functional dimensions. Self-report remains an insufficiently independent measure of such reorganization, motivating the search for convergent neural evidence. An analogous limitation, however, applies to much of existing neuroimaging literature on meditation and related practices: studies that measure neural activity during practice document state-dependent changes, and cannot by themselves establish whether the underlying organization persists outside of it. The present section, therefore, proceeds in two stages: first, examining whether neuroimaging findings converge on the structural dimensions defined in Section 3; then, asking whether any evidence exists for trait-level persistence of such reorganization beyond the context of practice. As in the preceding section, we refer to findings from these studies as bearing on awakening-like reorganization, since none of the cited work directly assessed the construct as operationalized here.

## 5.1 Neural Correlates of Self-Referential Processing

Neuroimaging studies of long-term meditators provide partial evidence across several of these dimensions, though findings on affective reactivity and temporal organization remain mixed and without clear convergence at the neural level, a pattern consistent with the classical traditions' explicit dissociation of awakening from affective and absorptive states (Berkovich-Ohana et al., 2013; Kral et al., 2018; Saketh et al., 2026). The studies reviewed here focus, therefore, on self-content coupling and self-referential processing, where convergence is strongest; although they differ in task design, modality, and participant characteristics, they point in a consistent functional direction rather than a single, unified neural mechanism.

Farb et al. (2007) demonstrated that self-referential processing is not unitary, distinguishing a narrative mode (temporal integration; medial prefrontal cortex, mPFC, and posterior cingulate cortex, PCC) from an experiential mode (present-moment processing; right insula/lateral PFC). In novices, these modes showed strong functional coupling, whereas in a cross-sectional sample of mindfulness-trained participants, they were decoupled. The neural systems supporting narrative self-reference and present-moment processing thus seem to dissociate, suggesting that self-related processing may be a modifiable feature of the cognitive system rather than a fixed constraint.

Brewer et al. (2011) provide converging evidence by showing that deactivation of the posterior cingulate cortex and medial prefrontal cortex generalized across three distinct meditation types in experienced meditators. Functional connectivity differences between meditators and controls were also present during the resting-state baseline, not only during active practice, suggesting a pattern associated with meditation experience rather than one generated only in the moment of practice. Josipovic et al. (2012) added a further specificity: nondual awareness meditation reduced the anti-correlation between intrinsic and extrinsic networks, whereas focused attention increased it, suggesting modulation of the coupling structure between internally and externally oriented processing.

Dor-Ziderman et al. (2013) observed in a neurophenomenological MEG study that selflessness does not simply correspond to a more complete version of narrative self-reduction but recruits a distinct beta-band network extending into posterior medial and lateral parietal regions; a secondary analysis linking these results to participants' first-person descriptions further associated the subset reporting loss of agency and ownership specifically with attenuation in the right inferior parietal lobule. A subsequent single-case study localized a similar beta-band reduction to the posterior medial cortex during graded self-boundary dissolution (Dor-Ziderman et al., 2016). Trautwein et al. (2024) provided the most rigorous test of this signature: in a preregistered MEG study with 46 long-term meditators, high-beta reductions in posterior medial cortex were specific to the subgroup reporting full dissolution of agency, first-person perspective, and self-location. These reductions correlated with phenomenological interview scores but not with classical self-report measures, consistent with the structural limitation of self-report identified in Section 4.

Winter et al. (2020) provided a phenomenologically vivid but methodologically limited illustration of these signatures co-occurring in a single case. During an EEG–fMRI session, one highly experienced meditator (>50,000 hours) spontaneously reported content-free awareness including clear, aware openness without thoughts, sensory content, or sense of self, accompanied by decreased alpha power, increased theta power, increased dorsal attention network connectivity, and reduced posterior Default Mode Network (DMN) connectivity. The reduction in posterior DMN connectivity parallels the pattern reviewed above; the single-case nature of the observation precludes generalization beyond this individual.

Together, these findings indicate that self-referential processing is not a unitary construct and provide converging evidence consistent with reorganization primarily along the dimensions of self–content coupling and self-referential evaluation, as defined in Section 3. Specifically, the decoupling of narrative and experiential modes, reduced default mode network engagement, and altered large-scale network interactions are consistent with diminished identification with mental content and reduced dominance of self-referential processing. By contrast, evidence for reorganization in affective reactivity and temporal organization of experience remains limited and less consistent at the neural level.

These findings are, however, derived from states occurring during or in close proximity to meditation practice, and most available evidence derives from paradigms probing default mode and large-scale network dynamics, which may bias the current picture. Whether this neural organization persists as a stable trait, independently of practice and context, remains the central unresolved question.

## 5.2 From State to Trait: Limits and Prospects

Bauer et al. (2019) provides the closest available fMRI evidence that network-level differences associated with meditation experience persist outside of practice, showing DMN reduction with increased Default Mode-Central Executive Network anti-correlation, the latter specific to meditators with under three years of practice, in the pre-meditation resting baseline of experienced meditators relative to controls. The study nonetheless examined long-term meditators rather than individuals whose awakening status had been independently assessed, and resting-state differences may therefore reflect cumulative practice effects rather than the structural reorganization Section 3 defines as awakening.

Complementary evidence comes from Lindström et al. (2023), who examined fMRI correlates of trait self-boundarylessness at rest and during self-referential processing in a mixed sample of meditators and non-meditators. mPFC activity during self-referential word processing showed an inverted U-shaped relation to boundarylessness, peaking at intermediate levels, and correlated with lifetime meditation experience, suggesting that self-boundary dissolution has a measurable neural correlate at the trait level, independent of active practice.

Extended cessation (EC), the volitional suspension of consciousness reported in advanced meditation, is a transient state, not itself a trait-level organization, though some traditions associate it with processes leading toward awakening. The evidence base for EC is still emerging: the studies cited below are, at the time of writing, available only as preprints. Yang et al. (2025) used 7T fMRI to characterize EC, finding increased unimodal activity alongside down-regulation of transmodal regions and DMN hubs; Shinozuka et al. (2025) provided concurrent EEG and MEG evidence of alpha reduction and increased neural complexity, distinguishing EC from sleep or anaesthesia. Converging with these findings, recent dense-sampling EEG microstate analyses further show that EC involves a systematic reconfiguration of large-scale brain dynamics, particularly affecting microstates associated with self-referential processing, alongside altered transition probabilities consistent with a precision re-weighting account of self-referential processing (Zarka et al., 2026).

These results suggest that EC does not merely suppress conscious content but transiently reshapes the dynamical balance between self-related and sensory-integrative processes. Its relevance here lies in the post-EC phenomenology: practitioners report heightened clarity and equanimity following re-emergence. Together with the downregulation of transmodal, DMN-associated regions observed during EC itself, this raises the possibility that what matters is not the cessation itself but a reorganization of self-referential processing that the cessation may trigger, precisely the kind of durable shift the present framework predicts.

By contrast, Hagerty et al. (2013) recorded, in a single case, EEG and fMRI during meditative absorption (jhana), finding nucleus accumbens activation and reward-system engagement alongside increased alpha power in regions associated with reduced external awareness. That a state of considerable meditative depth produces dopaminergic reward-system engagement rather than the network-level reorganization documented above illustrates that depth and intensity are not reliable proxies for awakening, a distinction the classical traditions reviewed in Section 2 make explicitly, and one the present neuroimaging evidence supports.

Taken together, the findings reviewed here converge on reproducible neural correlates of reduced self-referential processing during meditation-related states, with partial and indirect evidence that some of this organization persists outside of practice. Literature does not yet provide a stable, context-independent neural signature corresponding to awakening as the present framework defines it: a trait-level reorganization across four functional dimensions, independently of practice context. Bridging this gap requires theoretical frameworks capable of translating that structural claim into testable predictions about large-scale neural dynamics, to which we now turn.

# 6. Dereification as a Transformational Pivot

Building on this convergent evidence, we now turn from background synthesis to our own account of awakening as a trait-level transformation. On this view, awakening involves a shift in the global dynamical organization of cognition. It raises a fundamental question for cognitive science: What mechanisms can support enduring, trait-level transformations that operate independently of changes in experiential content? Addressing this question may require moving beyond models that privilege representational content, toward frameworks capable of capturing system-level reconfiguration, including changes in coupling, stability, and dynamical constraints governing cognitive processes.

One candidate mechanism through which such reorganization may occur is dereification: a metacognitive process through which the mind ceases to treat mental contents, thoughts, sensations, and perceptions, as solid, autonomous, and independent entities. Rather than accepting these contents as direct reflections of an objective reality, dereification enables the practitioner to perceive them as transient mental events. This concept, formally bridging experimental psychology and Eastern contemplative concepts such as *Sunyata* (emptiness), provides a framework for understanding the transition from a dualistic to a nondual cognitive mode.

The formalization of dereification owes much to the work of Lutz and colleagues (2015), who integrated it into a "phenomenological matrix" of mindfulness-related practices. Within this framework, dereification is defined as a dimension of metacognitive monitoring that systematically reduces "experiential coupling", that is, the degree to which a subject is immersed in or "fused" with the content of their thoughts. Building upon this, Dusana Dorjee (2016; Dorjee et al., 2025) proposed a Dereification Gradient, suggesting that this process is not binary but scales in depth. It begins with the monitoring of intrusive thoughts and progressively extends to the very structure of self and sensory perception, eventually leading to a profound shift in how reality is categorized and experienced.

The partial and indirect neural evidence reviewed in the preceding section is consistent with a gradient account of this kind: convergent findings of reduced self-referential processing during meditation-related states, alongside the absence of a stable, context-independent trait-level signature, are precisely what the dereification gradient predicts for populations in whom the process is ongoing rather than fully resolved.

## 6.1 How Dereification Shifts Experience Toward a Relational Mode

As dereification deepens, it can give rise to moments of *discontinuity* within the stream of consciousness. Under typical conditions, conscious experience is characterized by a strong sense of continuity and coherence, often accompanied by the implicit assumption of a stable self interacting with a stable world. Dereification disrupts this assumption by weakening the processes that maintain experiential solidity. This disruption can manifest phenomenologically as a break or gap in the usual flow of experience, a transient destabilization of what might be described as a "solid stream of consciousness." Such discontinuities do not necessarily imply pathology; rather, within contemplative contexts, they are often interpreted as revealing the constructed nature of experiential continuity itself. From a cognitive perspective, these discontinuities may reflect a temporary destabilization of predictive and integrative processes that ordinarily maintain experiential coherence across time. Crucially, the occurrence of discontinuity marks a transition point. It opens the possibility for a reorganization of experience along fundamentally different principles. Following such destabilization, experience may reorganize from a mode in which conventional distinctions are automatically reified toward a relational mode of perception. As dereification progresses, these distinction may no longer be perceived as inherently existing entities, but rather as emergent features of ongoing interactions. This shift gives rise to what can be described as a *relational mode of experience*, in which phenomena are understood primarily in terms of their dynamic interdependencies rather than their intrinsic properties. Importantly, this relational mode should not be confused with relativism; rather, it reflects a transformation in how experience is structured, moving from static representations to process-based interactions. Perception is no longer dominated by reified categorical and object-based processing, but becomes increasingly sensitive to context, dynamics, and co-emergence. This restructuring has important implications for both cognition and affect, as it reduces the rigidity associated with reified interpretations and allows for greater flexibility and nuance in experience.

## 6.2 Nondual Awareness as a Candidate State Within the Path to Awakening

At the apex of this gradient lies nondual awareness (NDA), representing the point of maximum dereification. In this state, the habitual distinction between the "observer" and the "observed object" collapses. Critically, however, pure NDA in its absolute form cannot constitute a stable endpoint for an organism embedded in physical and social world: the complete dissolution of self-object distinctions would undermine the cognitive architecture through which organisms orient, act, and coordinate with others. In such total absence of reification, the representational markers required for ego-centric action, such as distinguishing the biological self from an external object or prioritizing homeostatic needs, would be expected to dissolve, with phenomenological reports describing such conditions as "pure presence" (Damasio, 1999; Metzinger, 2004). Awakening, on the present account, is therefore not the permanent elimination of self-related processing but its reorganization within the cognitive system: a loosening of automatic identification that preserves functional engagement. This distinguishes the present account from the assumption that more ego dissolution is inherently better.

Northoff's World-Brain Relation theory provides potential neurobiological grounding for interpreting this reorganization. The theory posits that consciousness is grounded in the brain's intrinsic spatiotemporal dynamics rather than in isolated higher-order functions (Northoff et al., 2020). Time and space constitute a shared spatiotemporal format through which neural activity and experiential content are structured. From this perspective, processes such as dereification may reflect alterations in the spatiotemporal organization of intrinsic brain activity, leading to changes in how self-related and world-related content are integrated. Within the Topographic Reorganization Model of Meditation (Cooper et al., 2022), an extension of this account to advanced meditation practice, this network-level reorganization is framed not as a state-dependent effect but as a trait-level reconfiguration of the brain's intrinsic topographic structure, consistent with the account proposed here.

A more recent extension of this framework proposes that the brain's spatiotemporal organization, particularly the topographic centrality and dynamic timescales of the default mode network (DMN), corresponds to the subjective temporal structure of experience, including its expansion into past and future (Northoff et al., 2025). This offers a candidate mechanism through which dereification could plausibly extend to temporal organization of experience, though this remains a theoretical hypothesis rather than empirical evidence directly related to awakening.

Northoff's model potentially provides a framework to distinguish between contemplative and pathological states. While depression is characterized by abnormal organization of resting state activity, manifesting as increased self-focus and past-orientation (Northoff, 2016), and schizophrenia is characterized by disrupted temporal structure and altered DMN-Central Executive Network balance, resulting in a breakdown of the distinction between internally and externally directed mental contents (Northoff & Duncan, 2016), awakening may reflect a reorganization of intrinsic spatiotemporal dynamics consistent with the dereification gradient proposed here.

### 6.3 An Enactive Framework for the Operationalization of Awakening

The enactive approach to cognition posits that cognition is not the passive representation of a pre-given world, but rather the mutual "bringing-forth" of a world and a mind through dynamic coupling between organism and environment. Within this framework, experience is inherently relational, emerging from ongoing interactions rather than from independently existing constituents.

To synthesize the theoretical propositions put forward in this paper, we propose the EDM (see Figure 2), which operationalizes the transition from a reified to a relational mode of experience across training and awakening. At its core (Figure 2, gradient axis), the model posits that experience shifts from a mode of 'mutual identification', in which subject and object are reified as independent entities, toward a relational mode of 'co-emergence'. This reorganization is depicted, for the enactive structure of cognition, by nested spheres of context, body and self whose boundaries morph from rigid and separated to porous and mutually exchanging. Critically, the framework does not entail the permanent eradication of reification, but a reorganization of the system's dynamics: greater flexibility in moving between more reified, action-oriented engagement and more dereified, integrative modes. Finally, the distribution of momentary states (Figure 2, top) illustrates awakening as a trait-like shift in the relative time spent in dereified versus reified modes: by increasing the system's dynamical flexibility, the practitioner gains transient access to states of nondual awareness—without residing there and without compromising the adaptive function of self-related orientation. This transition from a fixed to a switchable mode of consciousness represents the transformational axis at the heart of the awakening process.

**Figure 2.**

*The enactive-dereification model (EDM) of awakening (conceptual schematic).*

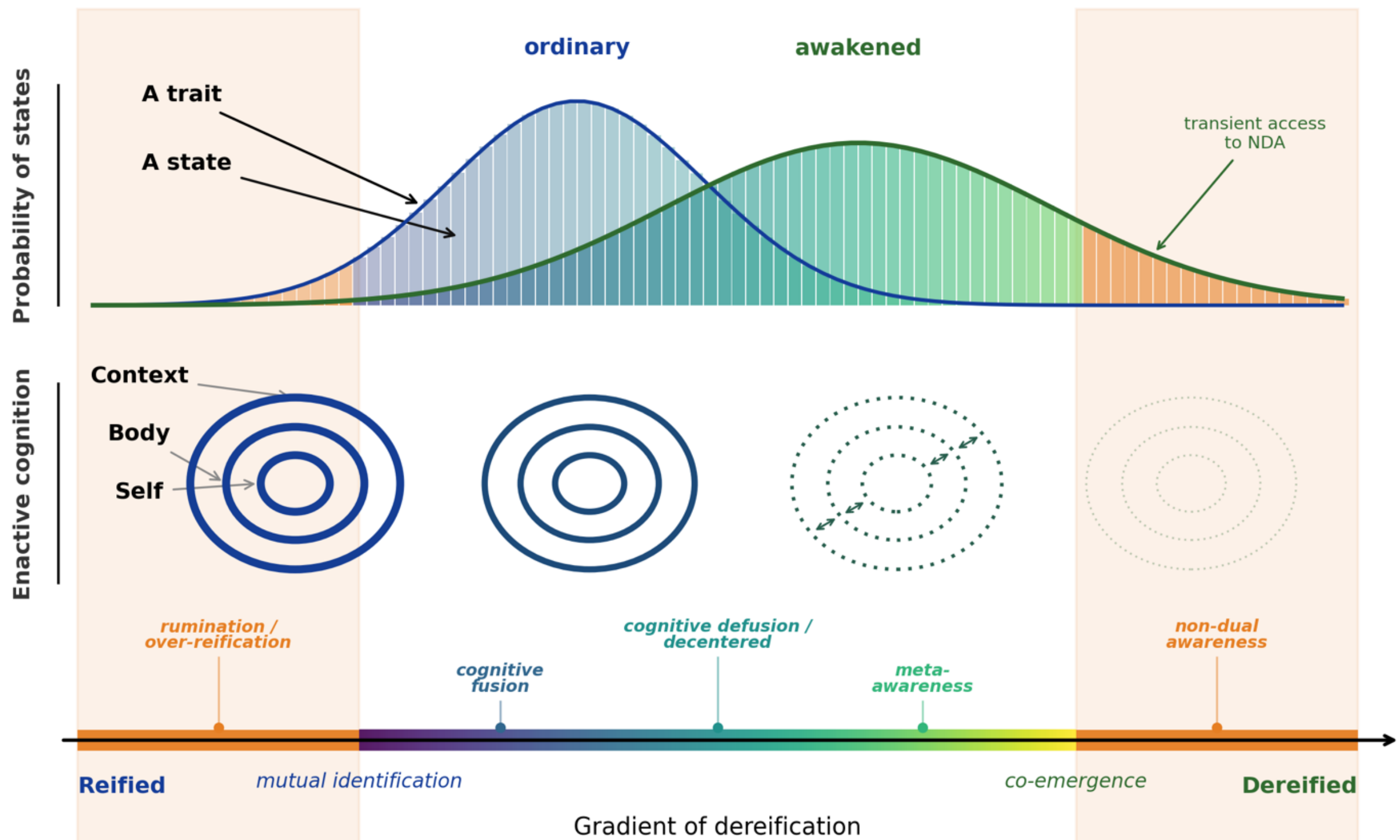

*Note.* The horizontal axis is the momentary state — the degree of dereification, from a reified pole (left) to a dereified pole (right); both extremes are non-functional and potentially pathological (orange). Top: over time an individual visits many states (each bar), whose distribution defines a trait (the time spent in each state). An ordinary trait is centred toward the reified pole and an awakened trait toward the dereified pole, the latter retaining only transient access to non-dual awareness (NDA) rather than residing at it. Bottom, the enactive-cognition paradigm: nested spheres — an outer context/environment ring, a middle body ring, and an inner self — whose boundaries morph from rigid and separated (reified mode; "mutual identification") to porous and mutually exchanging (dereified mode; "co-emergence"), depicting how dereification loosens the coupling among self, body, and world while preserving functional engagement.

From this perspective, dereification fundamentally transforms the nature of this co-emergence of subject and object. When the constituents of experience, such as the subject, perceptual objects, and conceptual structures, are no longer reified as stable and independent constituents, the resulting experiential field becomes less constrained by habitual schemas of grasping and aversion. In other words, altering how the elements of experience are structured as separate and fixed necessarily alters the structure of their interaction. As the constituents of experience lose their fixed, independent status, the dynamics of their co-emergence are reorganized. As dereification stabilizes over time, this relational mode may itself become stable, giving rise to a further degree of trait-level reorganization along the dereification gradient, characterized by increased flexibility, reduced automaticity, and a decreased tendency toward reified interpretation.

Table 1 summarizes this account across the four functional dimensions introduced in Section 3.4. For each dimension it contrasts the reified and dereified modes, notes the associated lived expression, and lists the convergent neural markers reviewed in Section 5, making explicit both where the evidence converges (self–content coupling and self-referential evaluation) and where it remains mixed or indirect (affective reactivity and temporal organization).

**Table 1.**

*Functional dimensions of awakening as a reified-to-dereified reorganization: the contrast between modes, the associated lived expression, and convergent neural markers.*

| **Functional dimension** | **Reified mode (ordinary)** | **Dereified mode (awakened)** | **Lived / behavioural expression** | **Convergent neural markers** |
|---|---|---|---|---|
| Self–content coupling (identification with mental contents) | Thoughts, emotions and perceptions automatically appropriated as 'me' or 'mine' and taken as direct reflections of reality | Contents experienced as transient mental events; reduced automatic appropriation; a decentred stance | Less fusion with thoughts; reduced rumination; greater psychological flexibility | Decoupling of narrative (mPFC/PCC) and experiential (insula) self-reference (Farb et al., 2007); PCC/mPFC deactivation across practices (Brewer et al., 2011); reduced posterior-DMN connectivity (Winter et al., 2020); right-IPL attenuation with loss of ownership (Dor-Ziderman et al., 2013) |
| Affective reactivity (perception-to-emotion escalation) | Perception tightly coupled to evaluative escalation; grasping and aversion | Weakened escalation; equanimity; non-reactive appraisal | Lower emotional reactivity; faster recovery; reduced craving and aversion | Mixed and indirect; classical traditions dissociate awakening from affective states (Berkovich-Ohana et al., 2013; Kral et al., 2018) |
| Temporal organization of experience (narrative projection across past and future) | Experience strongly narratively structured and projected across past and future | Present-centred; reduced narrative projection | Less mental time-travel; more present-moment engagement | DMN topographic centrality and timescales indexing subjective temporal structure (Northoff et al., 2025); contrasts with depression and schizophrenia (Northoff, 2016) |
| Self-referential evaluation (self-monitoring, commentary, judgment) | Continuous self-monitoring and judgment, stable and dominant | Loosened evaluative grip; reduced dominance of self-referential processing | Less self-criticism; reduced self-focus; steadier sense of self | Reduced mPFC/PCC self-referential processing (Farb et al., 2007; Brewer et al., 2011); trait-level DMN reduction at rest (Bauer et al., 2019) |

*Note.* The trait is characterized both by (i) a shift of the state distribution's mean toward the dereified pole and (ii) increased mobility around that mean; both extremes of the gradient are non-functional (see Figure 2).

The framework generates the following falsifiable hypotheses. Crucially, if dereification operates as proposed within the present framework, it should give rise to specific and testable empirical signatures. At the neural level, individuals reporting stable awakening-like transformations would be expected to exhibit reduced dwell time in self-related network configurations at rest, alongside increased flexibility in transitions across large-scale brain networks, potentially indexed by salience-network-mediated switching between internally and externally oriented processing (cf. Josipovic et al., 2012). Critically, these neural signatures should track independently assessed trait-level transformation rather than cumulative practice hours or generic meditation experience, a distinction consistent with the state-to-trait problem raised in Section 5.2. This pattern is consistent with emerging resting-state findings in long-term meditators (e.g., Panitz et al., 2025). In addition, intrinsic timescale organization within key regions such as the posterior medial cortex and DMN may be altered in this population, consistent with recent theoretical proposals linking default mode network spatiotemporal dynamics to the temporal structure of experience (Northoff et al., 2025; Wolff et al., 2022; Wolman et al., 2023), reflecting a reorganization of spatiotemporal integration that persists independently of active practice.

Importantly, these predictions are distinguishable from both general meditation effects and pathological conditions. Unlike transient meditation-induced states, which are expected to diminish outside of practice, the changes described here should persist at rest as trait-like alterations in cognitive dynamics. At the same time, they differ fundamentally from conditions such as depersonalization or dissociation, which are typically characterized by reduced self-related processing associated with instability and impaired integration rather than flexible reorganization. In the present framework, awakening is characterized not by a loss of structure, but by a reorganization toward greater dynamical flexibility and coherence. The framework would accordingly be weakened if the predicted neural differences were observed only during active practice, or if they tracked practice hours or generic meditation experience rather than independently assessed trait-level transformation.

# 7. Conclusion

We have argued that awakening is best understood not as a transient altered state but as a trait-level organization of the global structure of experience: a durable shift in how mental contents are appropriated, reacted to, and evaluated, rather than a change in the contents themselves. The present account, grounded in the enactive framework, reconceives awakening as a transformation in the relational dynamics of cognition. By linking phenomenological reports to candidate processes of co-emergence and dereification, it offers a path toward bridging subjective experience and objective measurement, opening the door to systematic empirical investigation.

Testing this framework will require moving beyond a single assessment modality. Structured phenomenological interviews, such as the micro-phenomenological approach developed by Petitmengin (2006, 2017), are one candidate of which promise Trautwein et al. (2024) illustrate directly. Their effects tracked interview-based rather than self-report measures of self-dissolution, addressing the structural limitation of self-report identified in Section 4. Experience sampling, which probes experience repeatedly in real time across everyday contexts rather than relying on a single retrospective report, would speak more directly to the proposed context-independence of the changes. Informant reports offer another useful source of evidence because they do not depend on the same self-referential processes under investigation. Ideally, these measures would converge with behavioral or task-based indices of the four dimensions defined in Section 3, allowing the framework to be tested across several partially independent sources of evidence rather than through any one measure alone. Such task-based indices still need to be developed, and to our knowledge, no existing assessment approach combines phenomenological, ecological, third-person, and behavioral evidence in the way required here. Future work should look into specifying and validating such an assessment stack as a necessary next step for testing the account proposed here.

Concretely, the four functional dimensions defined above yield a composite trait index—the mean of the dereification distribution—which, combined with an index of transition flexibility across states (its mobility), provides a minimal, two-part operationalization to be validated convergently against the neural and interview measures noted here. We present this scheme as illustrative rather than as a validated instrument.

## Declarations

**Funding.** This research did not receive any specific grant from funding agencies in the public, commercial, or not-for-profit sectors.

**Declaration of competing interest.** The authors declare that they have no known competing financial interests or personal relationships that could have appeared to influence the work reported in this paper.

**Data availability.** No data were used for the research described in the article.

**Declaration of generative AI and AI-assisted technologies in the writing process.** During the preparation of this work the author(s) used Chatgpt and Claude in order to improve the language and assist in drafting figure schematics ; after using this tool the author(s) reviewed and edited the content as needed and take full responsibility for the content of the publication.

# References

Alexander, C. (2025). What is 4E cognitive science? *Phenomenology and the Cognitive Sciences*, 1–26. https://doi.org/10.1007/S11097-025-10055-W

American Psychiatric Association. (2013). *Diagnostic and statistical manual of mental disorders* (5th ed.).

Anālayo, B. (2021). Awakening or Enlightenment? On the Significance of bodhi. *Mindfulness*, *12*(7), 1653–1658. https://doi.org/10.1007/S12671-021-01628-2

Bauer, C. C. C., Whitfield-Gabrieli, S., Díaz, J. L., Pasaye, E. H., & Barrios, F. A. (2019). From State-to-Trait Meditation: Reconfiguration of Central Executive and Default Mode Networks. *ENeuro*, *6*(6), ENEURO.0335-18.2019. https://doi.org/10.1523/ENEURO.0335-18.2019

Berkovich-Ohana, A., Dor-Ziderman, Y., Glicksohn, J., & Goldstein, A. (2013). Alterations in the sense of time, space, and body in the mindfulness-trained brain: a neurophenomenologically-guided MEG study. *Frontiers in Psychology*, *4*(DEC), 912. https://doi.org/10.3389/FPSYG.2013.00912

Bodhi, B. (Trans.). (2000). *The connected discourses of the Buddha: A translation of the Saṃyutta Nikāya*. Wisdom Publications.

Brewer, J. A., Worhunsky, P. D., Gray, J. R., Tang, Y. Y., Weber, J., & Kober, H. (2011). Meditation experience is associated with differences in default mode network activity and connectivity. *Proceedings of the National Academy of Sciences of the United States of America*, *108*(50), 20254–20259. https://doi.org/10.1073/PNAS.1112029108

Bronkhorst, J. (1993). *The two traditions of meditation in ancient India*. Motilal Banarsidass Publishing House.

Bryant, E. F. (2009). The Yoga Sūtras of Patañjali. *New York*.

Butlein, D. (2009). *The Impact of Spiritual Awakening on Psychotherapy*.

Carney, J. (2020). Thinking avant la lettre: A Review of 4E Cognition. *Evolutionary Studies in Imaginative Culture*, *4*(1), 77. https://doi.org/10.26613/ESIC/4.1.172

Conze, E. (2001). *Buddhist Wisdom: The Diamond Sutra and The Heart Sutra* (Random House/Vintage).

Cooper, A. C., Ventura, B., & Northoff, G. (2022). Beyond the veil of duality—topographic reorganization model of meditation. *Neuroscience of Consciousness*, *2022*(1), 1–22. https://doi.org/10.1093/NC/NIAC013

Corneille, J. S., & Luke, D. (2021). Spontaneous Spiritual Awakenings: Phenomenology, Altered States, Individual Differences, and Well-Being. *Frontiers in Psychology*, *12*, 720579.

Dahl, C. J., Lutz, A., & Davidson, R. J. (2015). Reconstructing and deconstructing the self: Cognitive mechanisms in meditation practice. *Trends in Cognitive Sciences*, *19*(9), 515–523. https://doi.org/10.1016/j.tics.2015.07.001

Damasio, A. (1999). The Feeling of What Happens: Body and Emotion in the Making of Consciousness. *The Journal of Speculative Philosophy*, *15*(4), 323–326. https://doi.org/10.1353/JSP.2001.0038

Davis, J. H., & Vago, D. R. (2013). Can enlightenment be traced to specific neural correlates, cognition, or behavior? No, and (a qualified) yes. *Frontiers in Psychology*, *4*(NOV). https://doi.org/10.3389/FPSYG.2013.00870

Deutsch, Eliot. (1969). *Advaita Vedānta : A philosophical reconstruction*. East_west Center Press. University of Hawaii.

Dor-Ziderman, Y., Ataria, Y., Fulder, S., Goldstein, A., & Berkovich-Ohana, A. (2016). Self-specific processing in the meditating brain: a MEG neurophenomenology study. *Neuroscience of Consciousness*, *2016*(1). https://doi.org/10.1093/NC/NIW019

Dor-Ziderman, Y., Berkovich-Ohana, A., Glicksohn, J., & Goldstein, A. (2013). Mindfulness-induced selflessness: a MEG neurophenomenological study. *Frontiers in Human Neuroscience*, *7*(SEP), 582. https://doi.org/10.3389/FNHUM.2013.00582

Dorjee, D. (2016). Defining contemplative science: The metacognitive self-regulatory capacity of the mind, context of meditation practice and modes of existential awareness. Frontiers in Psychology, 7, 1788. https://doi.org/10.3389/fpsyg.2016.01788

Dorjee, D., Nguyen, T., & Märtins, O. (2025). Inventory of Modes of Existential Awareness (IMEA): A new tool for assessing distinct self-transcendent and non-dual awareness states [Preprint]. OSF. https://osf.io/preprints/osf/smqeg_v2

Easwaran, Eknath. (2007). *The upanishads* (Vol. 2). Nilgiri Press.

Farb, N. A. S., Segal, Z. V., Mayberg, H., Bean, J., Mckeon, D., Fatima, Z., & Anderson, A. K. (2007). Attending to the present: mindfulness meditation reveals distinct neural modes of self-reference. *Social Cognitive and Affective Neuroscience*, *2*(4), 313–322. https://doi.org/10.1093/SCAN/NSM030

Feuerstein, G. (1979). *The Yoga-sūtra of Patañjali : A New Translation and Commentary*. Dawson.

Flood, G. D. (1996). *An introduction to Hinduism*. Cambridge University Press.

Fort, A. O. (1998). *Jīvanmukti in transformation: Embodied liberation in Advaita and Neo-Vedanta*. State University of New York Press.

Garab Dorje. (1996). *The golden letters: The three statements of Garab Dorje, the first teacher of Dzogchen, together with a commentary by Dza Patrul Rinpoche entitled "The special teaching of the wise and glorious king"* (J. M. Reynolds, Trans.; C. Namkhai Norbu, Foreword). Snow Lion Publications.

Hagerty, M. R., Isaacs, J., Brasington, L., Shupe, L., Fetz, E. E., & Cramer, S. C. (2013). Case Study of Ecstatic Meditation: fMRI and EEG Evidence of Self-Stimulating a Reward System. *Neural Plasticity*, *2013*(1), 653572. https://doi.org/10.1155/2013/653572

Hanley, A. W., Nakamura, Y., & Garland, E. L. (2018). The Nondual Awareness Dimensional Assessment (NADA): New Tools to Assess Nondual Traits and States of Consciousness Occurring Within and Beyond the Context of Meditation. *Psychological Assessment*, *30*(12), 1625. https://doi.org/10.1037/PAS0000615

Josipovic, Z., Dinstein, I., Weber, J., & Heeger, D. J. (2012). Influence of meditation on anti-correlated networks in the brain. *Frontiers in Human Neuroscience*, *5*(JANUARY 2012), 183. https://doi.org/10.3389/FNHUM.2011.00183

Josipovic, Z. (2014). Neural correlates of nondual awareness in meditation. *Annals of the New York Academy of Sciences*, *1307*(1), 9-18. https://doi.org/10.1111/nyas.12261

Kabat-Zinn, J. (2003). Mindfulness-based interventions in context: Past, present, and future. In *Clinical Psychology: Science and Practice* (Vol. 10, Number 2, pp. 144–156). https://doi.org/10.1093/clipsy/bpg016

Kelly, E. F., & Whicher, I. (2015). Patañjali's Yoga Sūtras and the Siddhis. *Beyond Physicalism: Toward Reconciliation of Science and Spirituality*, 315–348. https://psycnet.apa.org/record/2015-11603-009

Kilrea, K. A., Taylor, S., Bilodeau, C., Wittmann, M., Linares Gutiérrez, D., & Kübel, S. L. (2023). Measuring an Ongoing State of Wakefulness: The Development and Validation of the Inventory of Secular/Spiritual Wakefulness (WAKE). *Journal of Humanistic Psychology*. https://doi.org/10.1177/00221678231185891

Kral, T. R. A., Schuyler, B. S., Mumford, J. A., Rosenkranz, M. A., Lutz, A., & Davidson, R. J. (2018). Impact of short- and long-term mindfulness meditation training on amygdala reactivity to emotional stimuli. *NeuroImage*, *181*, 301. https://doi.org/10.1016/J.NEUROIMAGE.2018.07.013

Larson, G. James., & Bhattacharya, R. Shankar (1987). The Encyclopedia of Indian Philosophies, Volume 4 *Sāṃkhya : a dualist tradition inIndian philosophy*. Motilal Banarsidass.

Lindahl, J. R., Fisher, N. E., Cooper, D. J., Rosen, R. K., & Britton, W. B. (2017). The varieties of contemplative experience: A mixed-methods study of meditation-related challenges in Western Buddhists. *PloS One*, *12*(5). https://doi.org/10.1371/JOURNAL.PONE.0176239

Lindström, L., Goldin, P., Mårtensson, J., & Cardeña, E. (2023). Nonlinear brain correlates of trait self-boundarylessness. *Neuroscience of Consciousness*, *2023*(1), niad006. https://doi.org/10.1093/NC/NIAD006

Lutz, A., Slagter, H. A., Dunne, J. D., & Davidson, R. J. (2008). Attention regulation and monitoring in meditation. *Trends in Cognitive Sciences*, *12*. https://doi.org/10.1016/j.tics.2008.01.005

Lutz, A., Jha, A. P., Dunne, J. D., & Saron, C. D. (2015). Investigating the phenomenological matrix of mindfulness-related practices from a neurocognitive perspective. *American psychologist*, *70*(7), 632.

Martin, J. A. (2020). Clusters of Individuals Experiences form a Continuum of Persistant Non-Symbolic Experiences in Adults. *Social Psychology Commons, Social Psychology and Interaction Commons*, *8*.

Mayeda, S. (1992). *A thousand teachings : The Upadesasahasri of Sankara*. State University of New York Press.

Metzinger, Thomas. (2004). *Being no one : the self-model theory of subjectivity*.

Millière, R., Carhart-Harris, R. L., Roseman, L., Trautwein, F.-M., & Berkovich-Ohana, A. (2018). Psychedelics, Meditation, and Self-Consciousness. *Frontiers in Psychology*, *9*. https://doi.org/10.3389/fpsyg.2018.01475

Mills, P. J., Peterson, C. T., Pung, M. A., Patel, S., Weiss, L., Wilson, K. L., Doraiswamy, P. M., Martin, J. A., Tanzi, R. E., & Chopra, D. (2018). Change in Sense of Nondual Awareness and Spiritual Awakening in Response to a Multidimensional Well-Being Program. *Journal of Alternative and Complementary Medicine (New York, N.Y.)*, *24*(4), 343–351. https://doi.org/10.1089/ACM.2017.0160

Muktibodhananda, S. (2008). *Hatha Yoga Pradipika: Light on Hatha Yoga*. Yoga Publications Trust.

Newberg, A. B., & Waldman, M. R. (2018). A Neurotheological Approach to Spiritual Awakening. *International Journal of Transpersonal Studies*, *37*(2), 10. https://doi.org/https://doi.org/10.24972/ijts.2018.37.2.119

Newen, A., Gallagher, S., & De Bruin, L. (2018). 4E Cognition: Historical Roots, Key Concepts, and Central Issues. *The Oxford Handbook of 4E Cognition*, 3–16. https://doi.org/10.1093/OXFORDHB/9780198735410.013.1

Nikhilananda, S. (1949). *The Upanishads : Katha, Iśa, Kena, and Mundaka*. *Vol.1* Harper & Brothers Publishers, New York.

Nikhilananda, S. (1952). *The Upanishads: Śvetāśvatara, Praśna, and Māṇḍūkya with Gauḍapāda's Kārikā* (Vol. 2). Harper & Brothers, New York.

Northoff, G. (2016). Spatiotemporal psychopathology I: No rest for the brain's resting state activity in depression? Spatiotemporal psychopathology of depressive symptoms. *Journal of Affective Disorders*, *190*, 854–866. https://doi.org/10.1016/J.JAD.2015.05.007

Northoff, G., & Duncan, N. W. (2016). How do abnormalities in the brain's spontaneous activity translate into symptoms in schizophrenia? From an overview of resting state activity findings to a proposed spatiotemporal psychopathology. *Progress in Neurobiology*, *145–146*, 26–45. https://doi.org/10.1016/J.PNEUROBIO.2016.08.003

Northoff, G., Wainio-Theberge, S., & Evers, K. (2020). Is temporo-spatial dynamics the ‘common currency’ of brain and mind? In Quest of ‘Spatiotemporal Neuroscience’. *Physics of Life Reviews*, *33*, 34–54. https://doi.org/10.1016/J.PLREV.2019.05.002

Northoff, G., Buccellato, A., & Ventura, B. (2025). The Default Mode Network and inner time consciousness. Current Opinion in Behavioral Sciences, 63, 101524. https://doi.org/10.1016/j.cobeha.2025.101524

Panitz, D. Y., Mendelsohn, A., Cabral, J., & Berkovich-Ohana, A. (2025). Long-term mindfulness meditation increases occurrence of sensory and attention brain states. Frontiers in Human Neuroscience, 18, 1482353. https://doi.org/10.3389/fnhum.2024.1482353

Petitmengin, C. (2006). Describing one’s subjective experience in the second person: An interview method for the science of consciousness. *Phenomenology and the Cognitive Sciences*, *5*(3–4), 229–269. https://doi.org/10.1007/S11097-006-9022-2

Petitmengin, C. (2017). Enaction as a Lived Experience: Towards a Radical Neurophenomenology. *Constructivist Foundations 12(2): 139-147, 2017*, *12*(2), 139–147. http://constructivist.info/12/2/139

Phillips, S. (2009). *Yoga, karma, and rebirth : A brief history and philosophy*. Columbia University Press.

Radhakrishnan, S. (1953). *The Principal Upaniṣads*. George Allen and Unwin Ltd, London.

Rinpoche, T. (2002). *Essential Practice: Lectures on Kamalashila’s Stages of Meditation in the Middle Way School*. Snow Lion Publications.

Saketh, M., Sasidharan, A., Ventura, B., Venugopal, R., Bauer, C. C., Tewarie, P. K., ... & Northoff, G. (2026). Non-duality in brain and experience of advanced meditators—key role for intrinsic neural timescales. *Communications biology*. https://doi-org.acces.bibl.ulaval.ca/10.1038/s42003-026-10473-9

Śaṅkarācārya [Śaṅkara] (1921). *Vivekacūḍāmaṇi of Śrī Śaṅkarācārya* (Swami Madhavananda, Trans.). Advaita Ashrama.

Saraswati, S. S. (2013). *Four Chapters on Freedom : Commentary on Yoga Sutras of Patanjali*. Yoga Publications Trust.

Shinozuka, K., Yang, W. F. Z., Potash, R. M., Sparby, T., & Sacchet, M. D. (2025). Neuroelectrophysiological correlates of extended cessation of consciousness in advanced meditators: A multimodal EEG and MEG study. *BioRxiv*, 2025.09.19.677455. https://doi.org/10.1101/2025.09.19.677455

Sierra, M. (2009). *Depersonalization: A new look at a neglected syndrome*. Cambridge University Press.

Simeon, D., & Abugel, J. (2006). *Feeling unreal : Depersonalization disorder and the loss of the self*. Oxford University Press.

Taylor, S., & Egeto-Szabo, K. (2017). Exploring awakening experiences: A study of awakening experiences in terms of their triggers, characteristics, duration and after-effects. *Journal of Transpersonal Psychology*, (49(1)). https://psycnet.apa.org/record/2018-04040-004

Tempone-Wiltshire, J., & Matthews, · Floren. (2024). *Embodied Minds: An Embodied Cognitivist Understanding of Mindfulness in Public Health*. https://doi.org/10.1007/s12671-024-02423-5

Trautwein, F. M., Schweitzer, Y., Dor-Ziderman, Y., Nave, O., Ataria, Y., Fulder, S., & Berkovich-Ohana, A. (2024). Suspending the Embodied Self in Meditation Attenuates Beta Oscillations in the Posterior Medial Cortex. *The Journal of Neuroscience*, *44*(26), e1182232024. https://doi.org/10.1523/JNEUROSCI.1182-23.2024

Van Dam, N. T., van Vugt, M. K., Vago, D. R., Schmalzl, L., Saron, C. D., Olendzki, A., Meissner, T., Lazar, S. W., Kerr, C. E., Gorchov, J., Fox, K. C. R., Field, B. A., Britton, W. B., Brefczynski-Lewis, J. A., & Meyer, D. E. (2018). Mind the hype: A critical evaluation and prescriptive agenda for research on mindfulness and meditation. *Perspectives on Psychological Science, 13*(1), 36–61. https://doi/10.1177/1745691617709589

Varela, F. J., Thompson, E., & Rosch, E. (1991). *The embodied mind: Cognitive science and human experience. Cambridge, Mass: MIT Press.* (M. M. Press. Cambridge, Ed.).

Vasu, S. C. (Trans.). (1914). *The Siva Samhita*. The Panini Office.

Whicher, I. (1998). *The Integrity of the Yoga Darâsana*. A reconsideration of classical yoga. State University of New York Press.

Wilber, Ken., Engler, Jack., & Brown, D. P. (1986). *Transformations of consciousness : conventional and contemplative perspectives on development*. 356.

Winter, U., LeVan, P., Borghardt, T. L., Akin, B., Wittmann, M., Leyens, Y., & Schmidt, S. (2020). Content-Free Awareness: EEG-fcMRI Correlates of Consciousness as Such in an Expert Meditator. *Frontiers in Psychology*, *10*, 3064. https://doi.org/10.3389/FPSYG.2019.03064

Wolff, A., Berberian, N., Golesorkhi, M., Gomez-Pilar, J., Zilio, F., & Northoff, G. (2022). Intrinsic neural timescales: Temporal integration and segregation. Trends in Cognitive Sciences, 26(2), 159–173. https://doi.org/10.1016/j.tics.2021.11.007

Wolman, A., Çatal, Y., Wolff, A., Wainio-Theberge, S., Scalabrini, A., El Ahmadi, A., & Northoff, G. (2023). Intrinsic neural timescales mediate the cognitive bias of self – Temporal integration as key mechanism. NeuroImage, 268, 119896. https://doi.org/10.1016/j.neuroimage.2023.119896

Yang, W. F., Kadambi, A., Abellaneda-Pérez, K., Mackin, G., Beslic, I., Potash, R., ... & Sacchet, M. D. (2025). Endogenous suspension and reset of consciousness: 7T fMRI brain mapping of the extended cessation meditative endpoint. *bioRxiv*, 2025-09.

Zarka, D., Winson, Yang, F. Z., Rassat, A., Potash, R., Sparby, T., & Sacchet, M. D. (2026). EEG brain reconfiguration during meditation-induced extended cessation of consciousness: A dense-sampling multi-participant microstate study. *BioRxiv*, 2026.02.10.705005. https://doi.org/10.64898/2026.02.10.705005